\documentclass[12pt,preprint]{aastex}
\usepackage{color}
\usepackage{ulem}
\usepackage{epsfig}

\shorttitle{Maximum likelihood fitting of X-ray power density spectra}
\shortauthors{Didier Barret and Simon Vaughan}

\def\4u{4U1608-522}
\def\bwl{$b_w^{leakage}$}

\def\gsim{\mathrel{\hbox{\rlap{\hbox{\lower4pt\hbox{$\sim$}}}\hbox{$>$}}}}
\def\lsim{\mathrel{\hbox{\rlap{\hbox{\lower4pt\hbox{$\sim$}}}\hbox{$<$}}}}
\def\minchi2{min($\chi^2$)}

\begin{document}

\title{Maximum likelihood fitting of X-ray power density spectra: Application to high-frequency quasi-periodic oscillations from the neutron star X-ray binary \4u}

%
%
\author{Didier Barret\altaffilmark{1}}
\affil{Institut de Recherche en Astrophysique et Plan\'etologie \& Universit\'e de Toulouse (UPS), 9 avenue du Colonel Roche, BP 44346, 31028, Toulouse Cedex 4, France}

\email{didier.barret@irap.omp.eu}

\and

\author{Simon Vaughan}
\affil{X-Ray and Observational Astronomy Group, University of Leicester, Leicester, LE1 7RH, U.K.}


\begin{abstract}
High frequency quasi-periodic oscillations (QPOs) from weakly magnetized neutron stars display rapid frequency variability (second timescales) and high coherence with quality factors up to at least 200 at frequencies about 800-850 Hz. Their parameters have been estimated so far from standard \minchi2\ fitting techniques, after combining a large number of Power Density Spectra (PDS), as to have the powers normally distributed (the so-called Gaussian regime). Before combining PDS, different methods to minimize the effects of the frequency drift to the estimates of the QPO parameters have been proposed, but none of them relied on fitting the individual PDS. Accounting for the statistical properties of PDS, we apply a  maximum likelihood method to derive the QPO parameters in the non Gaussian regime. The method presented is general, easy to implement and can be applied to fitting individual PDS, several PDS simultaneously or their average, and is obviously not specific to the analysis of kHz QPO data. It applies to the analysis of any PDS optimized in frequency resolution and for low frequency variability or PDS containing features whose parameters vary on short timescales, as is the case for kHz QPOs. It is equivalent to the standard $\chi^2$ minimization fitting when the number of PDS fitted is large. The accuracy, reliability and superiority of the method is demonstrated with simulations of synthetic PDS, containing Lorentzian QPOs of known parameters. Accounting for the broadening of the QPO profile, due to the leakage of power inherent to windowed Fourier transforms, the maximum likelihood estimates of the QPO parameters are asymptotically unbiased, and have negligible bias when the QPO is reasonably well detected. By contrast, we show that the standard \minchi2\ fitting method gives biased parameters with larger uncertainties.

The maximum likelihood fitting method is applied to a subset of archival Rossi X-ray Timing Explorer (RXTE) data of the neutron star X-ray binary \4u, for which, we show that the lower kHz QPO parameters can be measured on timescales as short as 8 seconds. As to demonstrate the potential use of the results of the maximum likelihood method, we show that in the observation analyzed the time evolution of the frequency is consistent with a random walk. We then show that the broadening of the QPO due to the frequency drift scales as $\sqrt{T}$, as expected from a random walk ($T$ is the integration time of the PDS). This enables us to estimate the intrinsic quality factor of the QPO to be $\sim 260$, whereas previous analysis indicated a maximum value around 200.
\end{abstract}


%
%

\section{Introduction}
The standard method of weighted least squares (i.e. minimum $\chi^2$, hereafter \minchi2) model fitting is only equivalent to maximum likelihood estimation (MLE) of the model parameters when the data to be fitted are normally (Gaussian) distributed. As is well known, the distribution of $M$ averaged periodogram ordinates (from a stationary, linear stochastic process) follows a $\chi^2_{2M}$ distribution with $2M$ degrees of freedom; as $M$ increases, the $\chi^2_{2M}$ tends towards a normal distribution \citep{groth75, papadakis93}. In X-ray timing the most commonly used method to reach the so-called Gaussian regime is Bartlett's method \citep{bartlett48}: segment the original time series into $M$ non-overlapping segments, compute a periodogram for each segment, and average over all segments to produce a spectral estimate to be fitted \citep{vdk89}. Typically, one uses $M > 50$ to produce approximately Gaussian distributed averages \citep{papadakis93}. The drawbacks of this method are a loss of frequency resolution by a factor $M$ (reducing the sensitivity for narrow feature detection) and a suppression of the lowest frequencies. There is also a major drawback in the case of the analysis of high frequency quasi-periodic oscillations (QPOs), because their frequency varies rapidly with time, typically on timescales of seconds \citep{barret05b}. Since a kHz QPO can be a relatively narrow feature (FWHM $\sim 2-4$ Hz), sub-Hz frequency resolution of the PDS (equivalent to a segment duration of a few seconds) is required to have the QPO profile properly sampled. In order to use \minchi2\ fitting one must average a large number of segments, $M$, which then requires a total integration time exceeding hundreds, even thousands of seconds, depending of the strength of the QPO. On such timescale, the drift of the QPO frequency will smear out the QPO profile, leading to biases in the fitted QPO parameters, in particular the width. Methods to minimize the contribution of the frequency drift to the fitted QPO width have been proposed, using for instance the centroid (without fitting) of the excess power, as a best guess for the QPO frequency \citep{barret05b,barret06}. Although optimized, those methods suffer from the fact that statistical fluctuations of the noise can be stronger than the QPO signal, hence biasing the QPO frequency determination, especially when $M$ is small.

In this paper, we apply the maximum likelihood method to fit spectral models of high frequency QPOs in the non-Gaussian regime. To our knowledge, this is the first time such a method is applied to the analysis of {\it RXTE} PCA data, although it was discussed early on by \citet{stella94} for the analysis of continuum PDS recorded with {\it EXOSAT}. Such a method has been used already in a variety of applications, such as the modeling of solar oscillation spectra \citep{anderson90,appourchaux11} or fitting broad band PDS from long {\it XMM-Newton} observations of variable Seyfert 1 galaxies \citep{v10}. In section \ref{sec2}, we describe the Maximum Likelihood method. Then in section \ref{sec3}, the reliability and robustness of the technique is demonstrated using simulations of synthetic PDS. Its superiority over \minchi2\ fitting is illustrated in the case of $M=4$, and a comparison between the MLE and the \minchi2\ fitting is also presented for typical $M$ values used in QPO analysis. In section \ref{sec4}, the method is applied to a small subset of the archival {\it RXTE} data of the neutron star X-ray transient \4u, and the results of the MLE are used to infer the intrinsic width of the QPO.

%
%

\section{Description of the maximum likelihood method}
\label{sec2}
For describing the method, we follow the formalism of \cite{v05,v10}.  The periodogram of any linear stochastic process time series of length $N$, denoted $I_j = I(f_j)$ at Fourier frequency $f_j = j/N \Delta T$ (with $j=1,\ldots,N/2$), is exponentially distributed about the true spectral density $S_j = S(f_j)$ ($S_j$ is also the expectation value at $f_j$). 
\begin{equation}
\label{eqn:pdist}
  p(I_j | S_j) = \frac{1}{S_j} \exp( -I_j / S_j ),
\end{equation}
\citep[see e.g.][]{groth75, priestley81, leahy83, vdk89, percival93, bloomfield00}.
For simplicity, we assume even $N$. As pointed out by \citet{v05,v10}, this is valid only for Fourier frequencies $j=1, 2, \ldots, N/2-1$, as the power at the Nyquist frequency ($j=N/2$) follows a $\chi^2_1$ distribution with one degree of freedom (unlike powers at the other frequencies which follow the $\chi^2_2$ distribution).

Assuming a model $S(\theta)$, with parameters $\theta\equiv \{\theta_1, \theta_2, \ldots, \theta_L\}$, the joint probability density of observing $N-1$ periodogram points $I_j$, given the model values $\hat S_j$ ($j=1,N-1$, ignoring the Nyquist frequency) is: 
\begin{equation}
\mathcal{L} = \prod_{j=1}^{N-1} p(I_j | S_j)=\prod_{j=1}^{N-1} \frac{1}{S_j} \exp( -I_j / S_j ),
\end{equation}
where $\mathcal{L}$ is the likelihood, a function of $\theta$. This is sometimes known as the {\it Whittle} likelihood. {The $\chi_2^2$ distribution and hence the Whittle likelihood are the asymptotic expressions but are usually extremely good approximations for
the range of $N$ typically examined in these studies}. Maximizing the likelihood $\mathcal{L}$ is equivalent to minimizing $\mathcal{S}\equiv -2\ln \mathcal{L}$. $\mathcal{S}$ is then: 
\begin{equation}
 \mathcal{S}  =  -2 \ln \mathcal{L}
              =  2 \sum_{j=1}^{N-1} \left\{ \frac{I_j}{ S_j} + \ln S_j \right\}.
\end{equation}

The above formula can in fact be generalized to the case of fitting the average of $M$ PDS, as discussed in Appendix A.

Finding the model parameters that minimize $\mathcal{S}$ yields the maximum likelihood estimates (MLEs) of the model parameter, $\hat\theta$ \citep{anderson90}. The minimization can be achieved with a standard numerical optimisation algorithm, such as {\it POWELL} or {\it AMOEBA} \citep[e.g.][chapter 10]{press92}.

It is easy to show that if the spectral model is a constant, $S(f) = a$, then its MLE, $\hat a$, can be found by minimising
\begin{equation}
\mathcal{S}=2 \left\{ (N-1)\ln a+1/a\sum_{j=1}^{N-1} I_j \right\}.
\end{equation}
The solution is found where the derivative of $\mathcal{S}$ against $a$ is null:
\begin{equation}
  \hat{a} = \frac{1}{N-1} \sum_{j=1}^{N-1} I_j.
\end{equation}
The MLE of the constant is simply the sample mean.

Confidence intervals can be computed from the so-called {\it Fisher matrix (F)}, the expectation value of the Hessian \citep[e.g.][and references therein]{andrae,heavens}. The error on $\hat \theta_i, \sigma_i $ is then computed as: 
 \begin{equation}
\sigma_i^2 = (F)^{-1}_{ii},~ F_{ij}=\langle-\frac{\delta^2 \ln \mathcal{L}}{\delta \theta_i\delta \theta_j}\rangle
\end{equation}

 The {\it Fisher matrix} can be evaluated numerically. The confidence limits on the MLEs can also be calculated from $\Delta \mathcal{S} = \mathcal{S}(\theta) - \mathcal{S}(\hat \theta)$ in the same manner as the popular $\Delta \chi^2$ method \citep{cash79}, e.g. $\Delta \mathcal{S}=1$ corresponds to the $68.3$\% confidence limits on one parameter \citep[see discussion in][]{v05}. The  Fisher matrix method is fast but gives reliable confidence intervals only when the likelihood surface is Gaussian; the $\Delta \mathcal{S}$ method is slower but gives reasonable confidence intervals even for non-Gaussian likelihoods and can be used to search for local minima. We have verified through simulations and with real data that the errors computed by the two methods did not differ by more than a few \%. 

One can also use standard tools of maximum likelihood analysis, such as the likelihood ratio test to test for additional free parameters in the model \citep[see][for the conditions of use]{v10,protassov02}. Unlike the $\min(\chi^2)$ method, the fit statistic used in the maximum likelihood (i.e. $\mathcal{S}$) does not provide an automatic goodness-of-fit test. However, as discussed in \citet{v10}, useful diagnostic statistics can be constructed and calibrated using Monte Carlo simulations that are sensitive to data-model mismatch.

The minimisation process can be made more robust against  local minima by (i) repeating the optimisation from several (random) settings of the initial parameter estimates, and (ii) monitoring the $\Delta \mathcal{S}$ values during the confidence interval calculations. Non-negative parameters, like $w$ and $R$ can also be logarithmically transformed\footnote{As discussed by \citet[][and references therein]{appourchaux11} these parameters tend to be lognormally distributed, hence the log-transformed parameters are approximately normally distributed.} to improve the reliability of the numerical methods \citep{anderson90} [equation \ref{lorentzian} below must then be rewritten with $R \rightarrow \exp(\ln R)$ and $w \rightarrow \exp(\ln w)$]. Immediately after fitting, $R$ and $w$ can then be transformed back in linear space for further processing.  With IDL, we used the {\it POWELL} minimization routine, but checked that the downhill simplex method implemented in the {\it AMOEBA} routine gave exactly similar results. A completely independent set of codes, from the generation of PDS to the MLE, was also developed in R (\url{http://www.r-project.org/}) on a different platform and yielded very similar and fully consistent results, with the IDL ones.
%
%

\section{Results}
\label{sec3}
We now wish to test the reliability of the MLE with simulated data containing a QPO of known and realistic parameters. For this, we need to define a model for the PDS. This model is the sum of a constant $a$ (to account for the Poisson noise, i.e. equal to $2$ for Leahy normalized PDS, \cite{leahy83}), plus a Lorentzian with three parameters, the normalization $R$, the width $w$ (FWHM) and the centroid frequency $\nu_0$.
\begin{equation}
S(\nu ; \theta) = a + \frac{R w}{2\pi ((\nu-\nu_0)^2+(w/2)^2)}
\label{lorentzian}
\end{equation}
(where we have denoted the parameters $\theta = \{a, R, w, \nu_0\}$).

\begin{figure}[!t]
\epsscale{0.6}
\plotone{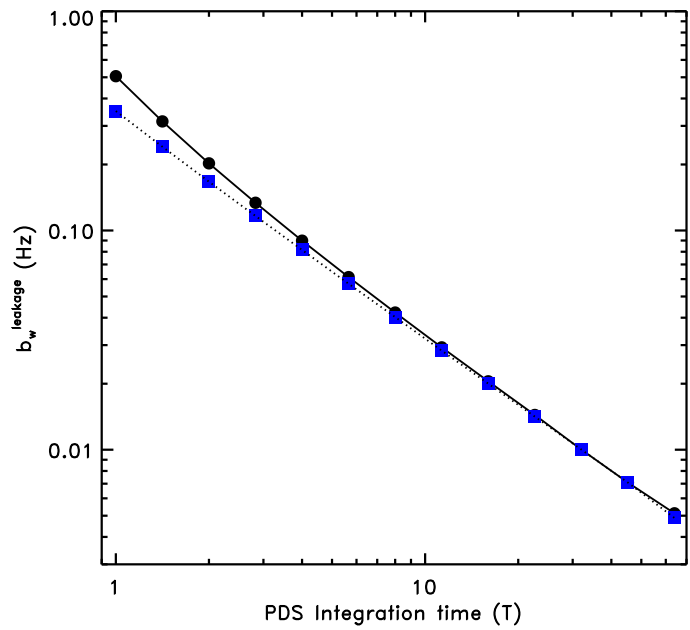}
\caption{Magnitude of the bias on $w$ (\bwl ) as a function of time series duration $T$ evaluated for two different values of $w$ (1 and 5 Hz; lower and upper curves respectively). The leakage bias was computed numerically by comparing of the Lorentzian profile before and after convolution (computed on a fine grid of frequencies, $\delta f=2^{-13}$). Roughly, the bias decays as \bwl $\propto T^{-1}$.\label{bv11r_f8}}
\end{figure}

Random time series are generated from the PDS model by inverse Fourier transforming the randomised data \citep{timmer95}. The time series are then processed to produce PDS of the required frequency resolution. As with the real data, the simulated time series should be much longer than the required PDS integration time. This method includes the well known effect of the windowing-induced power leakage, e.g. narrow features are slightly broadened \citep[see e.g.][]{uttley03, v10} (see Appendix B). This effect is quite significant for narrow features and short integration times, as shown in Figure \ref{bv11r_f8}. The bias on the width of the QPO caused by leakage (hereafter \bwl) goes approximately like $T^{-1}$. Beside this bias, we have checked that our simulations did not generate any biases on the other parameters, $R, \nu_0$ and $a$. This was done by averaging the periodograms from a large number of simulations, as an approximation to the expectation of the periodogram (Appendix B), and fitted this (using \minchi2 or maximum likelihood) to obtain the parameters of the expected periodogram. Within errors, the best fit parameters, denoted $\bar{R}, \bar{\nu_0}$, $\bar{a}$ and $\bar{w}$ corrected for the leakage bias were consistent with the input values used for the simulations. 

 \subsection{Biases in the MLEs}
In order to test the reliability of the method, we generated and fitted a large number of random datasets and computed the bias on the MLEs as the difference between the sample mean of the MLE estimates $<\theta_i^{MLE}>$ and the best fitted value of the average periodogram, such that for parameter $i$, we have $b_i = < \theta_i^{MLE} >-\bar{\theta_i}$. Alternatively, the bias can be estimated from the true value put in the simulations ($\theta_i^0$), so that we have  $b_i = < \theta_i^{MLE} > - \theta_i^{0}$. In this case, special care must be taken for computing the bias on $w$, as the measured value must be corrected for the leakage bias (\bwl), such that $b_w = < w^{MLE} > -b_w^{leakage} - w^0$. In this paper, we have used the first method to compute the bias. The MLEs are asymptotically unbiased,  but when estimated from a finite data set, some may be biased. This is a reason why the analysis of real data (especially when attempting to fit low signal to noise ratio features) should be compared with simulated data sets, as closely representative of the data as possible. However, as we will show below, some MLEs are unbiased, and for the biased ones, the bias can be made negligible (i.e. much smaller than the scatter of the data), when the MLE is applied to QPOs that are reasonably well detected (see discussion below for detection criteria).

\begin{figure}[!h]
\epsscale{0.6}
\centerline{\psfig{figure=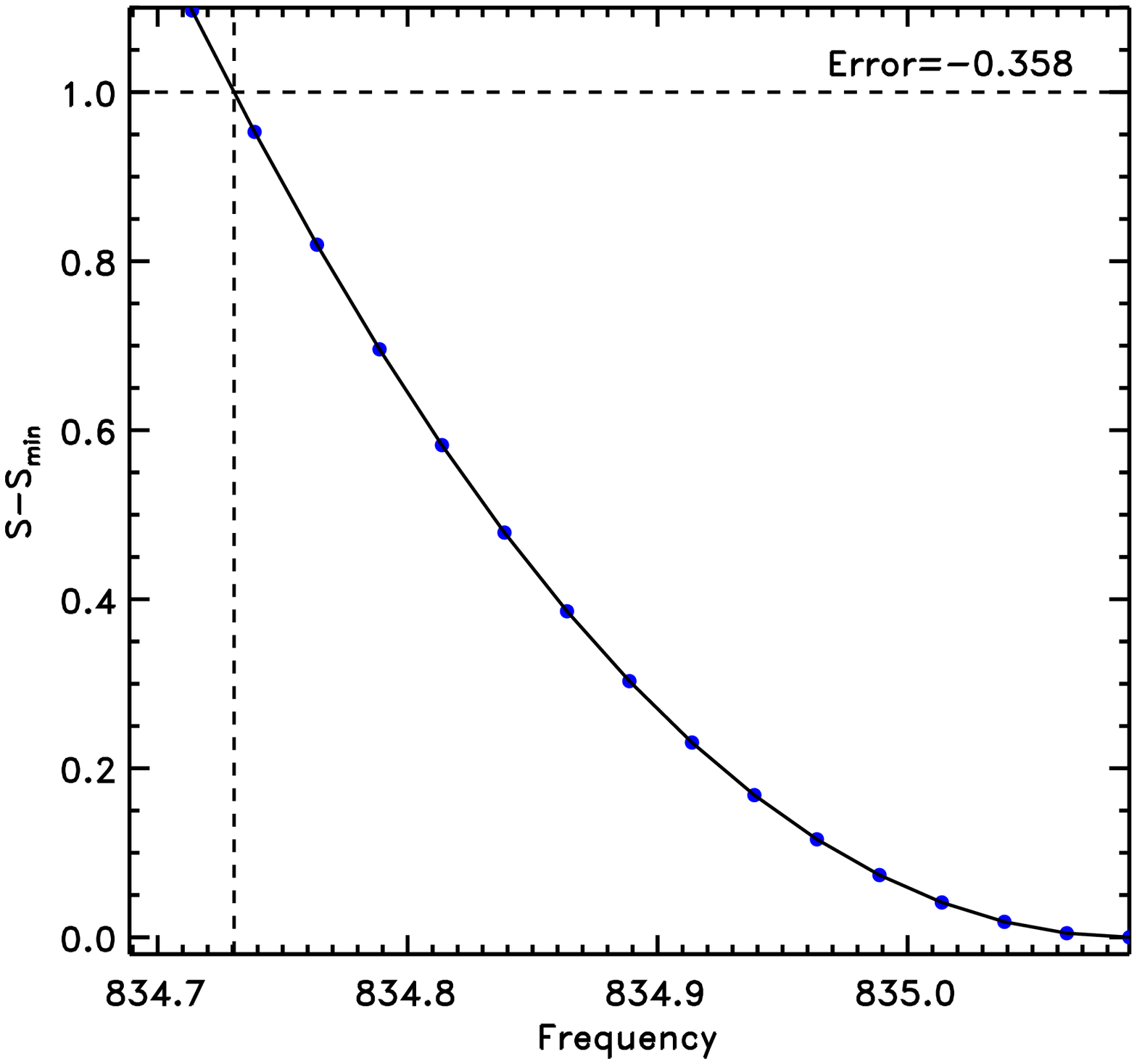,height=5cm,angle=0}\psfig{figure=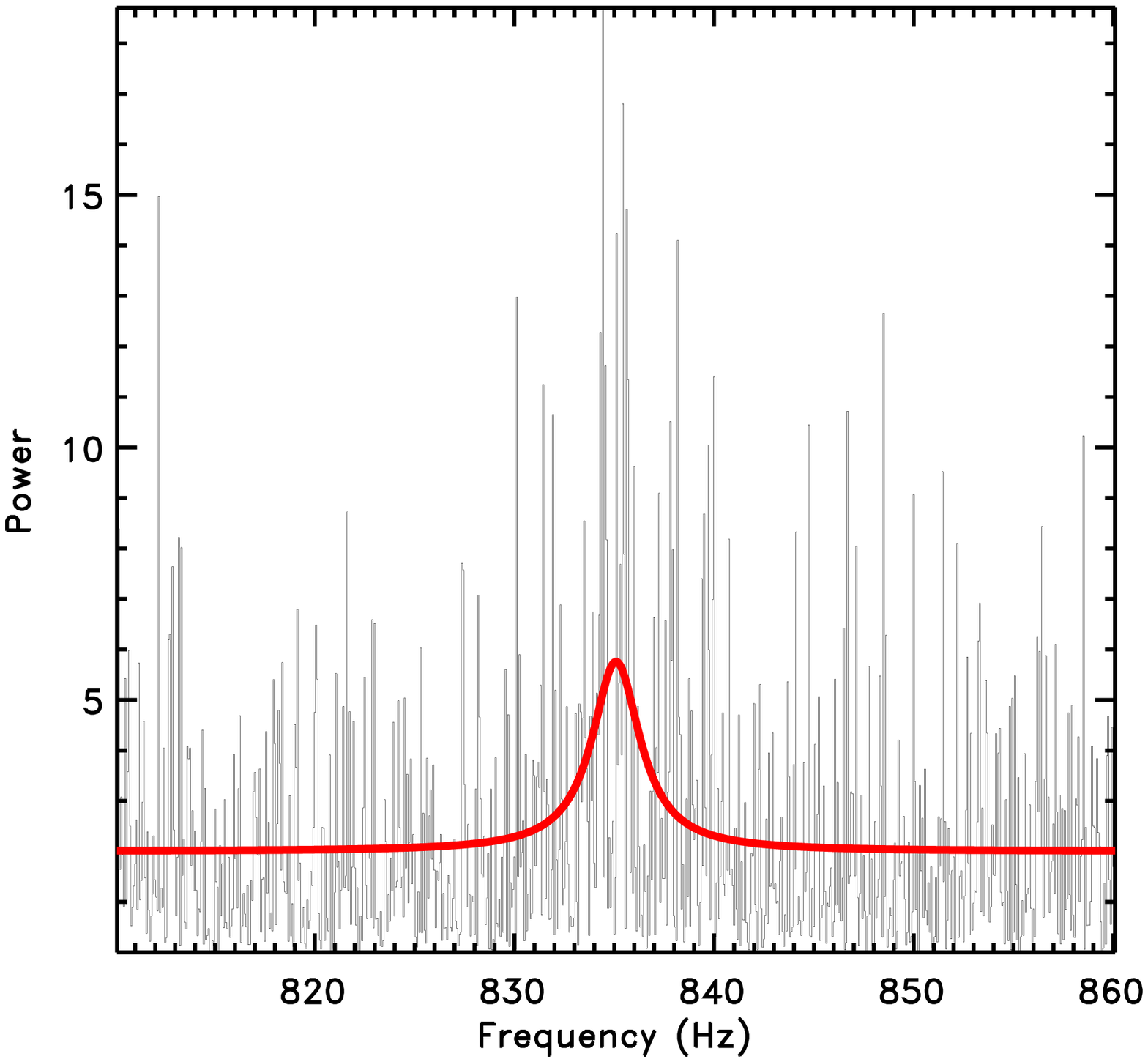,height=5cm,angle=0}\psfig{figure=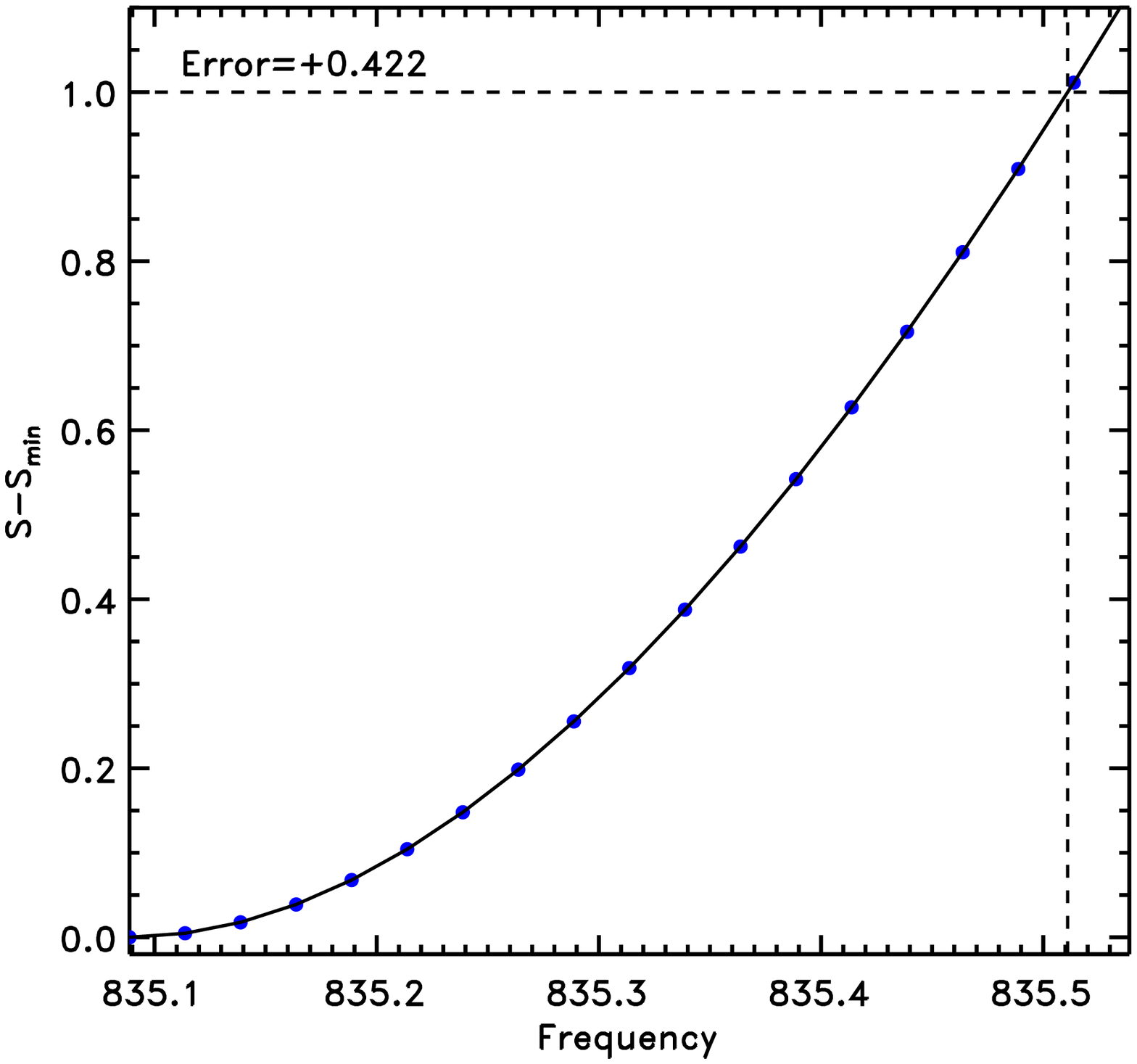,height=5cm,angle=0}}
\caption{MLE fit of one single 16 second PDS (center panel), together with the $\Delta \mathcal{S}$ curves for the estimate of the QPO frequency (left and right panels). Despite a very noisy PDS, the MLE picks up the QPO at the right frequency (835.08 Hz assumed in the simulation). The $1\sigma$ error on the QPO frequency is only about 0.4 Hz. \label{bv11r_f1}}
\end{figure}

As we are primarily interested in QPOs of a few Hz breadth, that can be detected and fitted on a few tens of seconds, we have generated periodograms with different integration times, corresponding to sub-Hz frequency resolutions (from $T= 8$ to $T=48$ seconds, in step of 4 seconds). For each $T$, 16384 PDS (128 sets of 128 PDS) were generated from light curves $64 \times T$ long, and fitted. A large number of simulations is required to give precise Monte Carlo estimates of the bias and variance of the MLEs. We have explored a wide range of model parameters, but here we have assumed $R=19.0, w=3.23$ Hz, $\nu_0=835.08$ Hz. $R=19.0$ corresponds to a QPO RMS amplitude of about $\sim 10$\% for a source count rate of $2000$ counts/s. Those parameters are consistent with the ones derived from the analysis of the \4u\ data presented in section \ref{sec4}. One example of an MLE for one single PDS (T=16 seconds) is shown in Figure \ref{bv11r_f1}. To increase speed, yet to enable a good determination of the Poisson level, the fits were performed on a frequency range of $250$ Hz on each side of the QPO peak. As can be seen, despite a very noisy periodogram, the maximum likelihood fitting picks up the QPO at the right frequency, with a $1\sigma$ uncertainty of only about 0.4 Hz.

\begin{figure}[!t]
\epsscale{1.125}
\plottwo{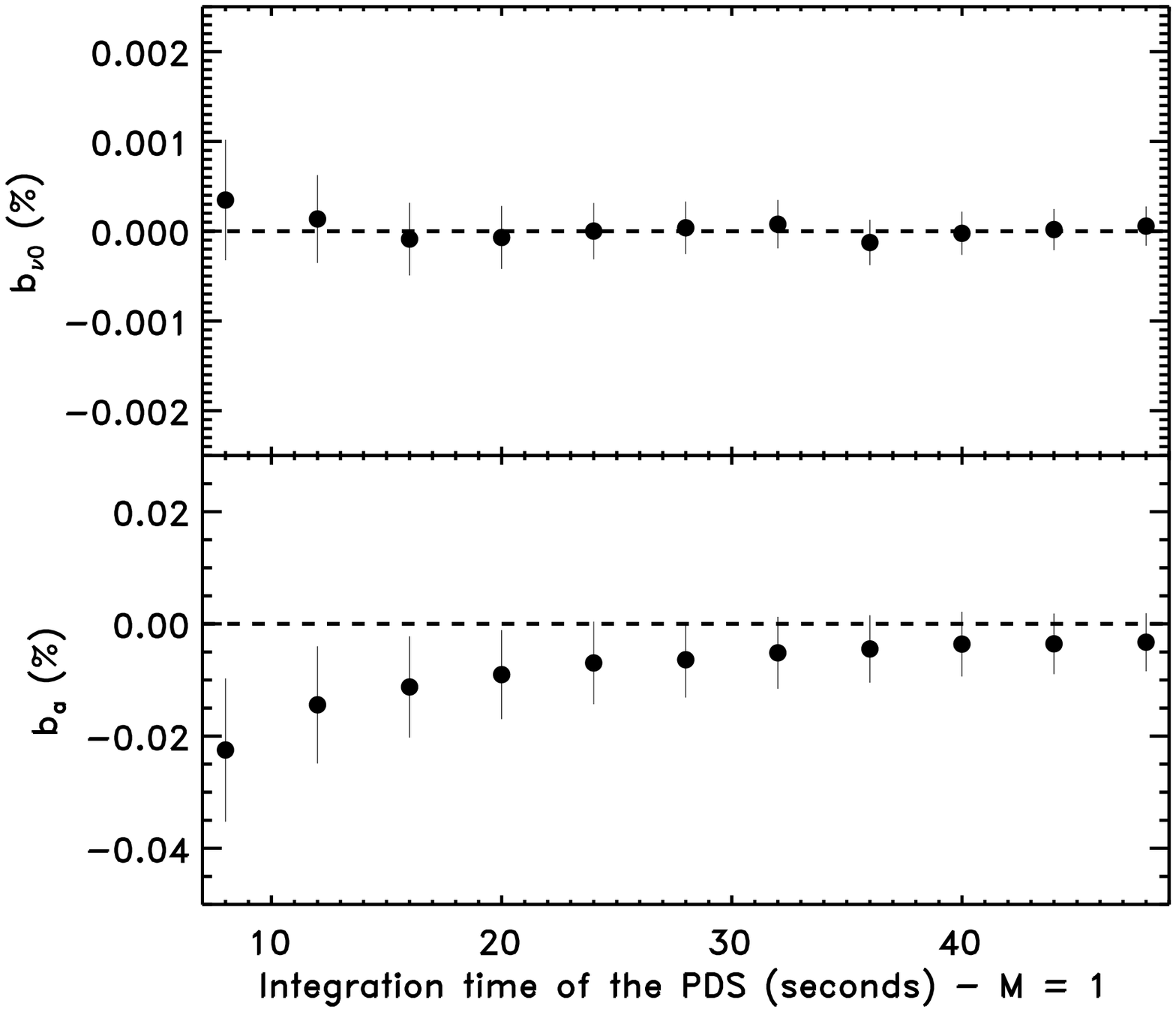}{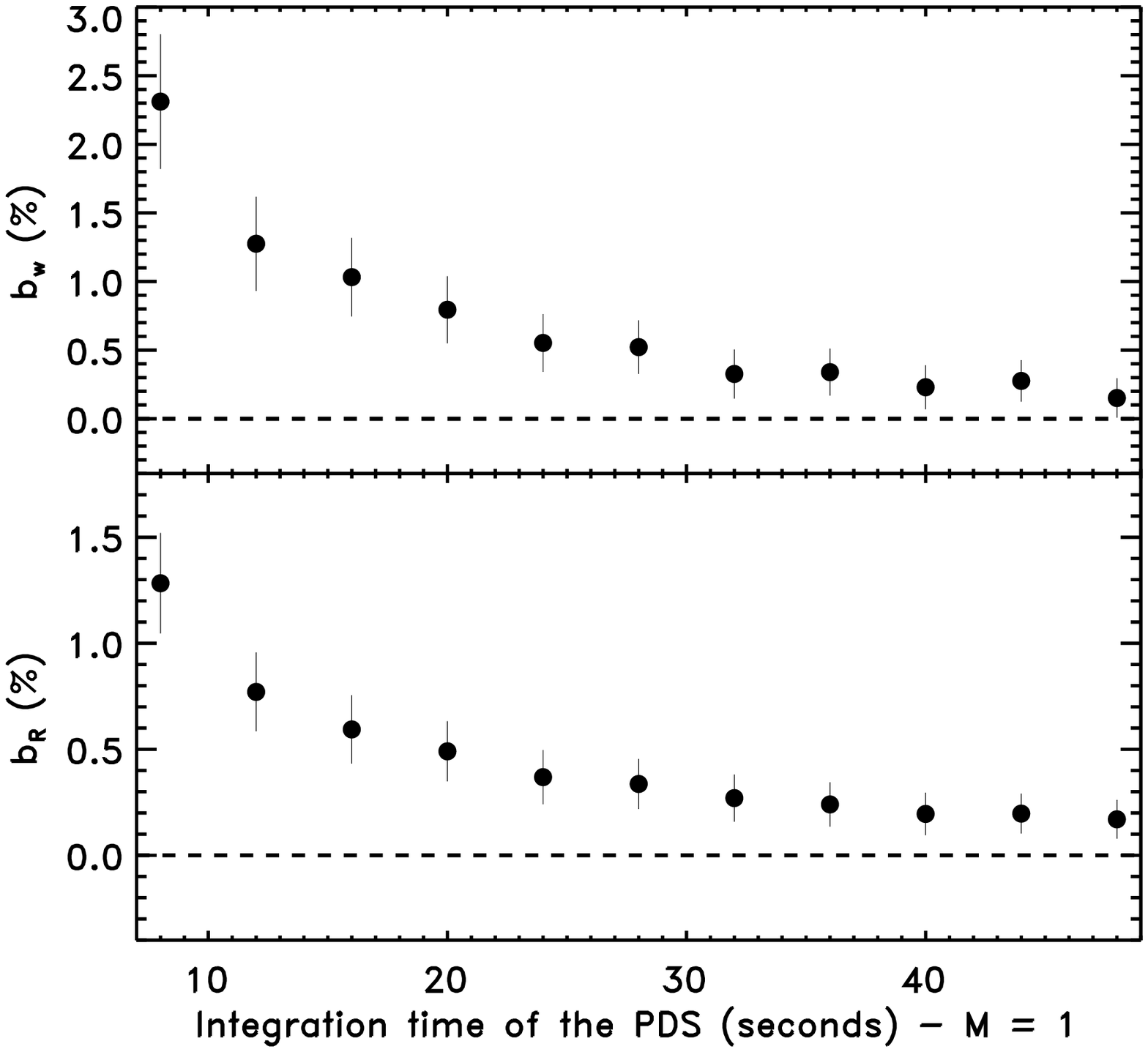}
\caption{Measured biases on the four model parameters (top left, $\nu_0$ , bottom left, $a$, top right, $w$, bottom right, $R$) as a function of the PDS integration time ($T$). Those biases are derived from the difference between the sample mean of the MLE estimates from single PDS ($M=1$) and the best fitted value of the average of the $16384$ PDS generated for each integration times, and are expressed in percentage of the best fitted averages. Error bars on the bias are derived from the standard error on the MLEs, they do not account for error on the fits to the averaged periodograms (which are generated from the same data and so are not independent of the averaged MLEs). As can be seen there is no bias in the MLEs of the QPO frequency and Poisson level, but there is a $\lsim 2-3$\%, hence negligible bias, in $R$ and $w$ at short PDS integration times (corresponding to a lower signal-to-noise ratio for the QPO). Note that 3\% corresponds to a bias of $\sim 0.1$ Hz on $w$ for the QPO considered here. As expected, these biases decrease when $T$ increases. \label{bv11r_f2}}
\end{figure}

The results on the biases are shown in Figure \ref{bv11r_f2}. As can be seen, the biases on the fitted Poisson level and on the QPO frequency are almost zero. There is a small remaining positive, but {\it negligible} bias on $R$ and $w$ that decreases with increasing $T$. The bias on $R$ and $w$ is much smaller than the typical $1\sigma$ of the estimates, e.g. by more than a factor of 10 for all $T$, hence can only be measured using a very large number of PDS, as is the case in our simulations ($N=16384$). As expected, the biases on $R$ and $w$ are more severe when the signal-to-noise ratio of the QPO decreases (e.g. for PDS of a given $T$ for a QPO of smaller $R$ or for PDS of shorter $T$ for a QPO at a given $R$, see Figure \ref{bv11r_f2}). \citet{boutelier09} considered a significant QPO detection when the ratio between the fitted $R$ and its $1\sigma$ error is larger than 3 or equivalently when the mean significance of the excess power fitted\footnote{This is defined as $n_\sigma=\frac{\bar{P}-P_{mean}}{2/\sqrt{MW}}$, with $\bar{P}$ the mean of MW powers, $P_{mean}$ is the Poisson level (close to 2), M the number of averaged PDS, W the number of frequency bins averaged, see \cite{boirin00}. A scanning algorithm maximizes $n_\sigma$ with respect to $W$, allowing $W$ to vary within a plausible range, say 2 to 10 Hz, for narrow QPOs.} exceeds $ 6\sigma$. In Figure \ref{bv11r_f2}, both criteria are met, even for $T=8$ seconds, for which $<R/\sigma_R>=3.5$ and $<n_\sigma>=9.2$. This means that, provided that $T$ is adjusted as to ensure that QPOs are reasonably well detected (i.e. one of the two criteria above is met), the MLE will not introduce any significant biases on the estimates of the QPO parameters. For lower signal-to-noise ratio QPOs, as stressed above, it is recommended to quantify the residual biases through simulations representative of the data set analyzed. 

\begin{figure}[!t]
\epsscale{1.125}
\plottwo{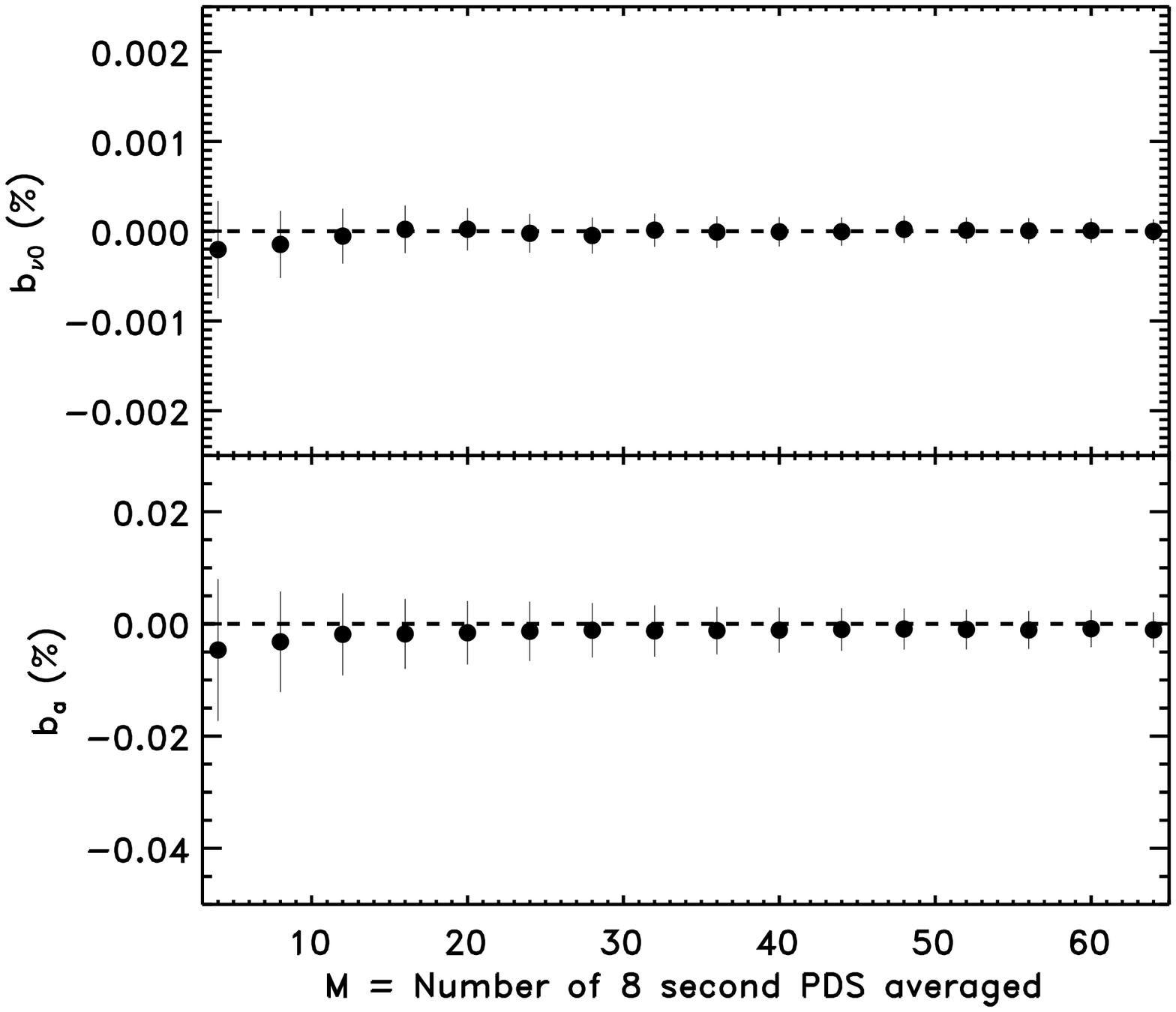}{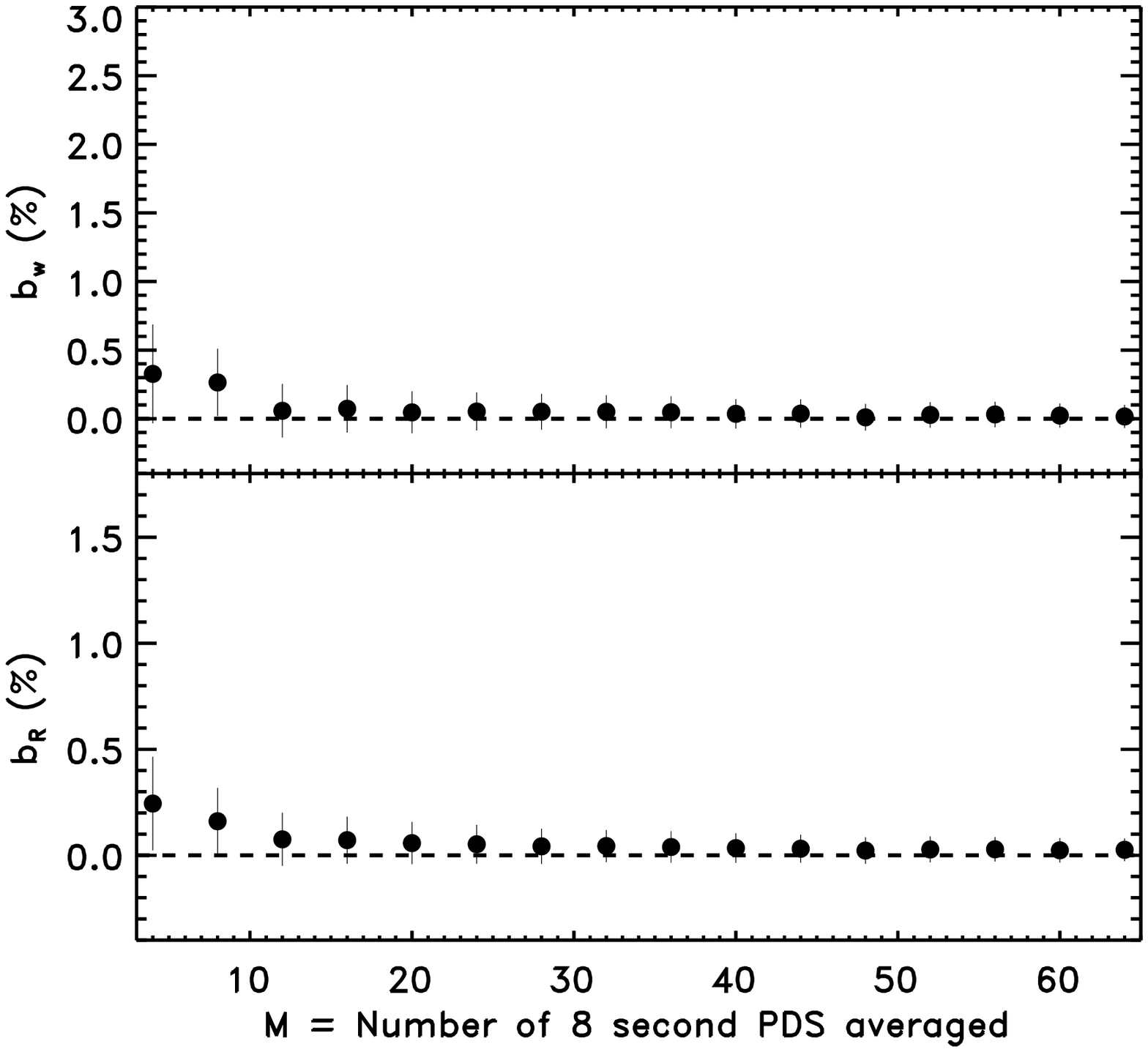}
\caption{Measured biases on the four model parameters (top left, $\nu_0$, bottom left, $a$, top right, $w$, bottom right, $R$), as a function of the number of 8 second PDS averaged (same y-axis scale as figure \ref{bv11r_f2}). Those biases are derived from the difference between the sample mean of 4096 MLEs (each one from fitting the average of M PDS), and an estimate of the average QPO parameters generated by the simulation over the entire set of $4096\times M$ PDS simulated. The biases are expressed in percentage of these average values. Error bars on the bias are derived from the standard error on the MLEs, they do not account for error on the fits to the averaged periodograms (which are generated from the same data and so are not independent of the averaged MLEs). Clearly, all the biases converge to zero. \label{bv11r_f3}}
\end{figure}

The biases on the MLEs decrease with increasing $T$, as shown in Figure \ref{bv11r_f2}. We have extended the previous simulations to longer equivalent PDS integration times, adopting a different, though consistent, approach. We have generated $4096M$ simulations of $T=8$ second PDS ($M$ varying from 4 to 64). The range of $M$ considered correspond to PDS integration times that are typical of the QPO analysis performed so far \citep{barret05a}. The 4096 averages of the $M$ PDS were fitted by maximum likelihood (32 different sets of $128 \times M$ PDS), following Appendix A. The biases were again computed as the difference between the sample mean of the 4096 MLEs and the sample mean of 32 MLEs derived from fitting the average $128 \times M$ PDS, as an estimate of the mean QPO parameters over the entire $4096 \times M$ PDS simulated. The residual biases are shown in Figure \ref{bv11r_f3}. First, to illustrate the equivalence of fitting the average of $M$ 8 second PDS and fitting one single PDS of $M\times 8$ second integration time, it can be seen that the biases on $w, R$ (about 0.2-0.3 \%) measured for $M=4$ ($T=8$ seconds) are fully consistent with the biases measured for $M=1$ and $T=32$ seconds (see Figure \ref{bv11r_f2}). Second, as expected, all the biases converge smoothly to zero as $M$ increases.

\subsection{Comparison with \minchi2\ fitting}
\begin{figure}[!t]
\centerline{\psfig{figure=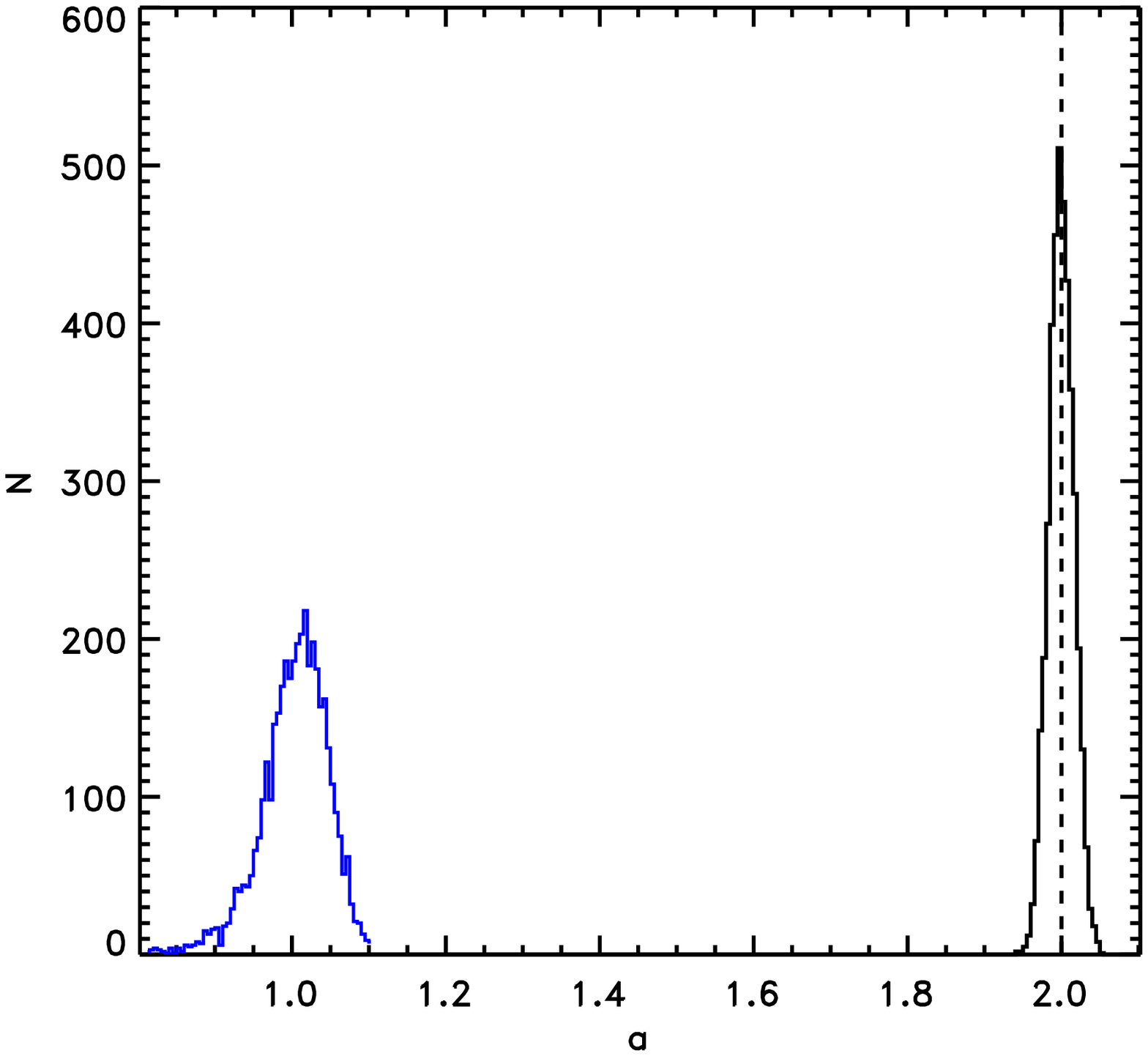,height=7.5cm,angle=0}\psfig{figure=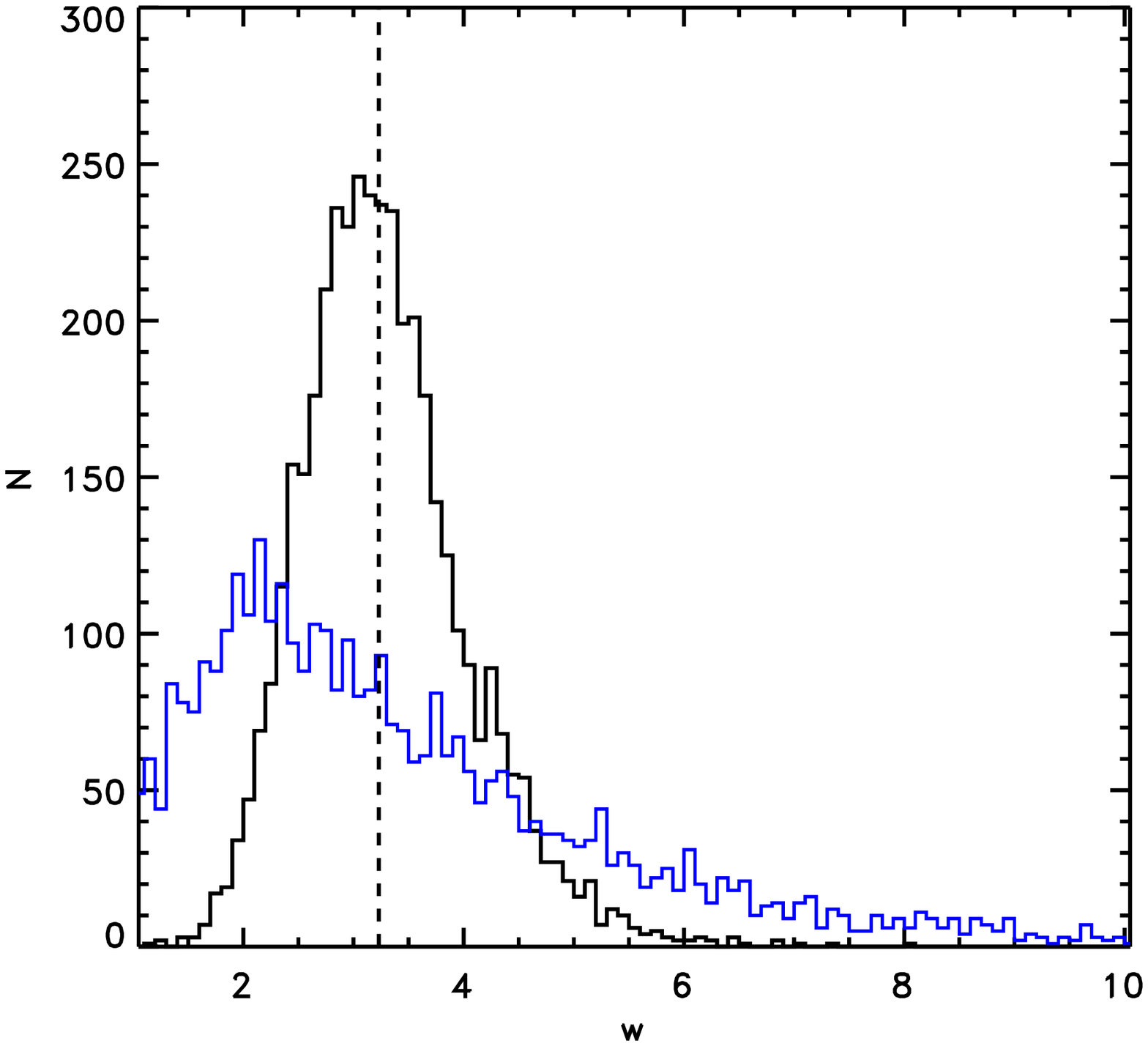,height=7.5cm,angle=0}}
\centerline{\psfig{figure=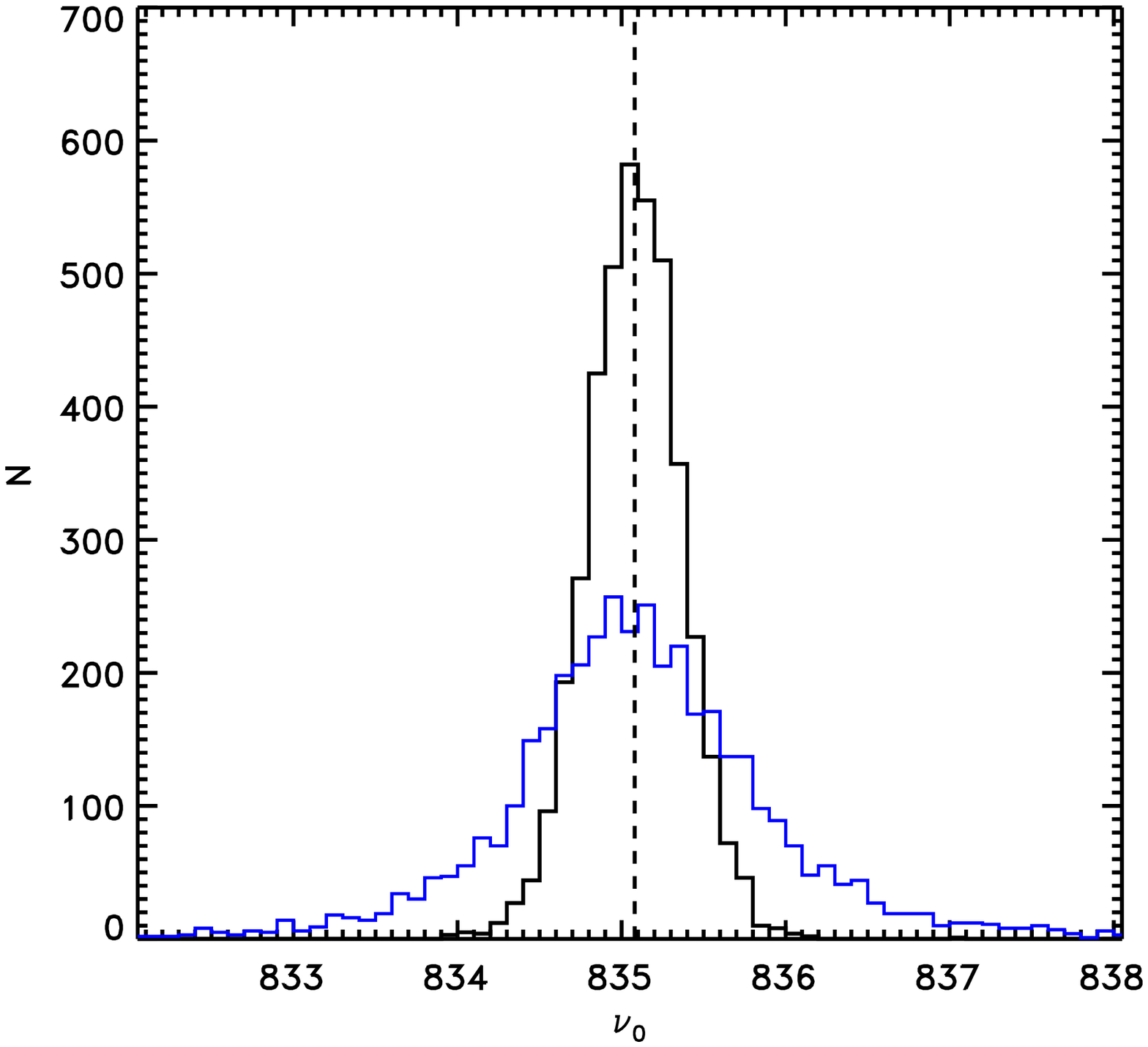,height=7.5cm,angle=0}\psfig{figure=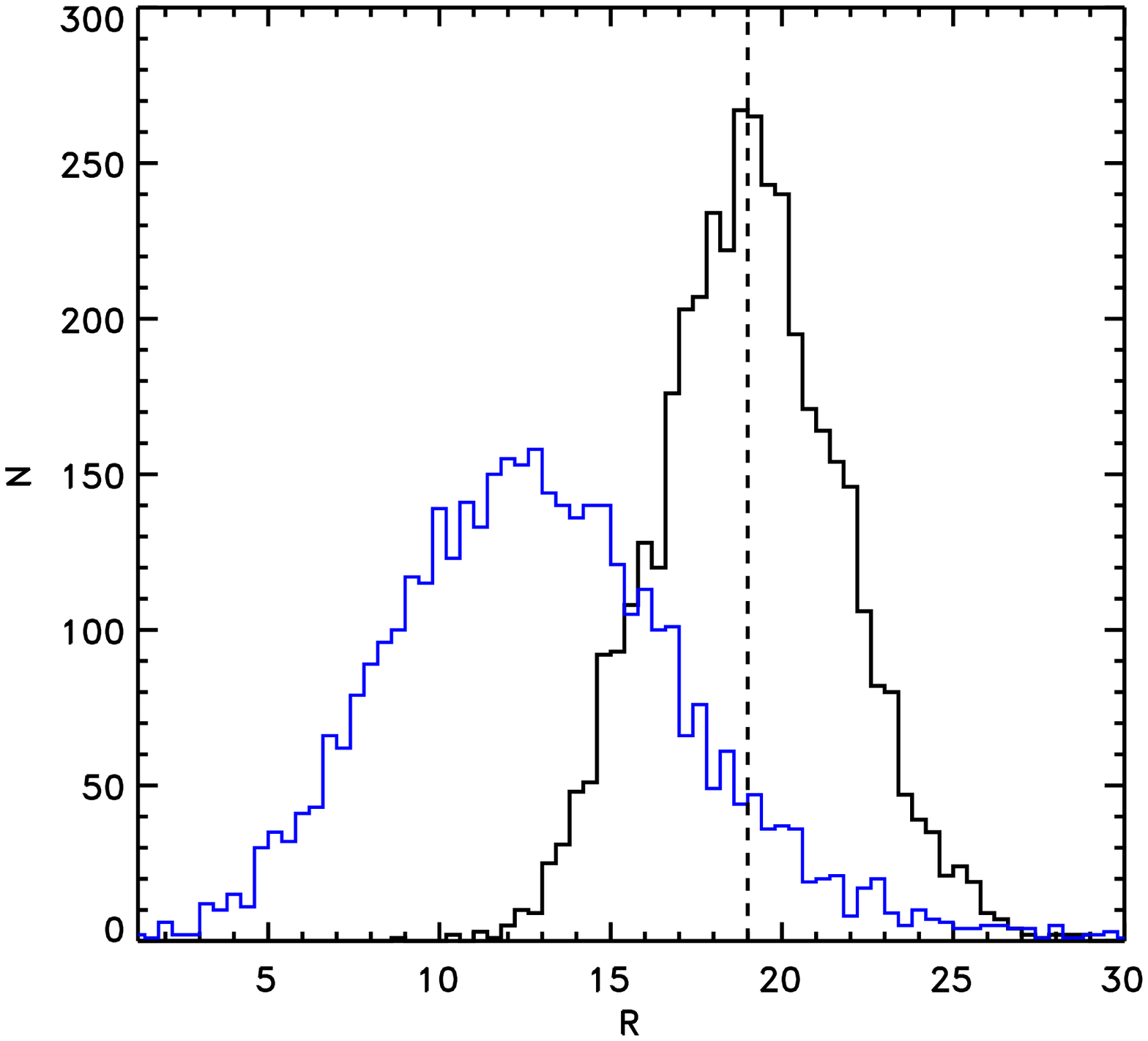,height=7.5cm,angle=0}}
\caption{Results of fitting simulated data with a  constant and a Lorentzian QPO model. Each of the four panels shows the histogram of the estimates for one of the four model parameters (top left, $a$, bottom left, $\nu_0$, top right, $w$, bottom right, $R$). The parameters were estimated using the MLE (black line) and the \minchi2\ method (blue line) by fitting the average of $M=4$ periodograms (4096 samples). For \minchi2\ fitting, outliers in the parameter distribution (e.g. fitted $R\ge 100$) were removed before building the histograms (about 2\% of the total sample are outliers). The vertical dashed line indicates the true parameter values for the simulations. Clearly the MLEs show no strong biases, but the \minchi2\ results underestimate both the Poisson level and the QPO amplitude. The estimates have generally larger spreads from \minchi2\ fitting than from the MLE. \label{bv11r_f4}}
\end{figure}

We have fitted simulated data with the MLE and \minchi2\ for comparison, using the 8 second PDS generated for the bias analysis discussed above. For illustrative purpose, the averages of $M=4$ PDS were fitted by both techniques (with the errors on each power $I_j$ set to $I_j/\sqrt{M}$ for \minchi2\ fitting). We chose $M=4$ because it corresponds to a timescale (32 seconds) on which the properties of  kHz QPOs have been investigated previously \cite[e.g.][for 4U1608-522]{barret05b}. To ensure a fair comparison of the MLE and \minchi2\ methods the same starting values were used for minimisation. The histograms of the estimates of each of the four model parameters, as derived from the ensemble of fits for the two methods are shown in Figure \ref{bv11r_f4}. The strongest biases in the \minchi2\ fitting clearly concern the Poisson level and the QPO amplitude. This is explained by the fact that the \minchi2\ distribution is asymmetric, producing more data below than above the expected spectrum, and these low powers, which have smaller errors, drive the fitted parameters down. By contrast the QPO frequency estimates obtained by \minchi2\ fitting are not significantly biased but do show a larger spread (i.e. larger uncertainty in the estimates) than the corresponding MLEs (the same applies to the QPO width). This shows the power of the MLE over \minchi2\ when fitting a small number of PDS. 

\begin{figure}[!t]
\epsscale{.60}
\plotone{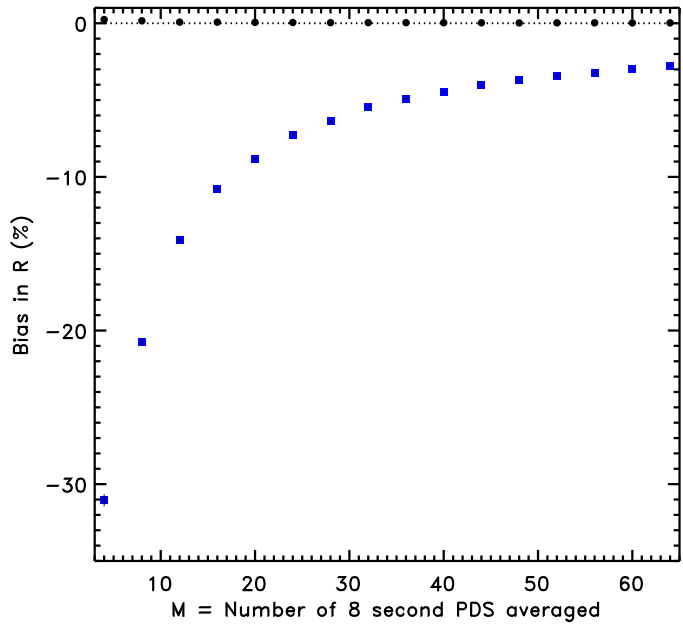}
\caption{Biases of fitted QPO amplitude ($R$) with \minchi2\ (blue filled squares) and MLE (black filled circles) fitting for different $M$, from 4 to 64, in steps of 4. Error bars on the MLEs are plotted but are in many cases smaller than the plot symbols. 4096 averages of $M$ PDS were fitted by the two methods. The bias of the MLEs of $R$ is lower than 0.3\% ($M=4$). At the opposite, biases of the $R$ estimates from  \minchi2\ fitting can be as large as $\sim 30$\% for $M=4$ and remains as large as $\sim 2.5$\% for $M=64$. This figure alone demonstrates that MLE should always be preferred, when averaging a very large number of PDS is not possible. \label{bv11r_f5}}
\end{figure}

As can be seen from Figure \ref{bv11r_f4}, there is quite a large bias in $R$ with \minchi2\ fitting, and this bias is important as $R$ gives the RMS amplitude of the QPO, one of the QPO parameter that is often reported in the literature. It is interesting to see how it decreases when $M$ increases. In the limit of a large $M$, the results from \minchi2\ fitting should be unbiased and consistent with the ones derived from the MLE. \cite{papadakis93} showed that for $M>50$ the powers of the PDS are approximately normally distributed, hence  \minchi2\ fitting can be used. Using the same simulated data set as used to generate Fig \ref{bv11r_f3}, i.e. $4096\times M$ PDS, we show how the bias on $R$ varies with $M$ with the two fitting techniques (see Figure \ref{bv11r_f5}). As expected the bias decreases with increasing $M$ with \minchi2\ fitting. It is however interesting to note that for $M=64$, the \minchi2\ fitting leads to a bias of about 2.5\%. This bias is larger than the one derived from the MLE when $M=1$ ! (see Figure \ref{bv11r_f2} for $T=8$ seconds).  With the MLE, the bias in the estimate of $R$ decreases with $T$, and is negligible for all $M$ (it is always below $\sim 0.3$\%). This means that whenever, averaging a very large number of PDS is not possible or desirable, MLE fitting should be preferred against \minchi2\ fitting, especially if one is interested to measure the QPO amplitude as accurately as possible.

\section{Application to RXTE/PCA data: the case of 4U1608--522}
\label{sec4}
\begin{figure}[!t]
\epsscale{1.1250}
\plottwo{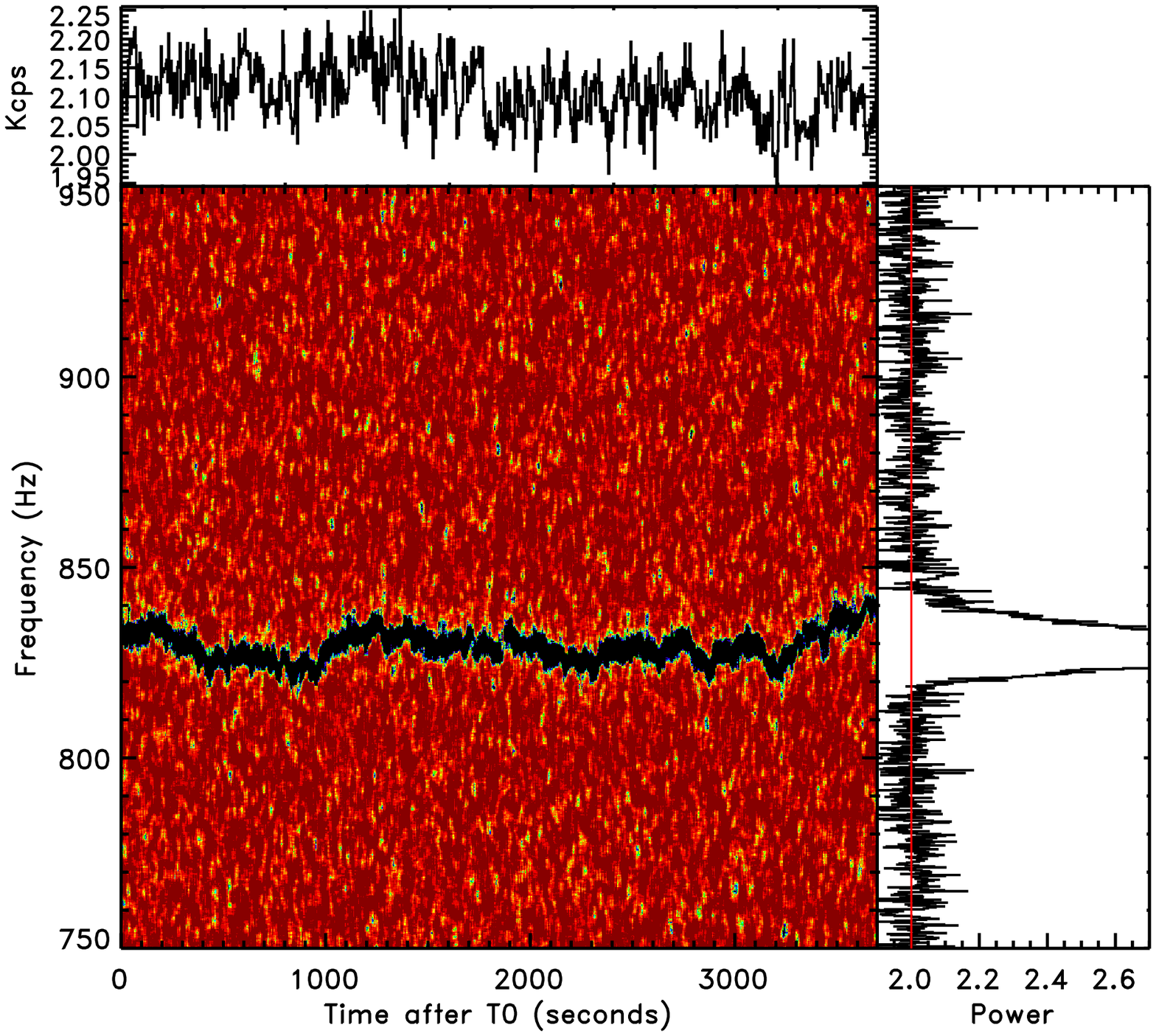}{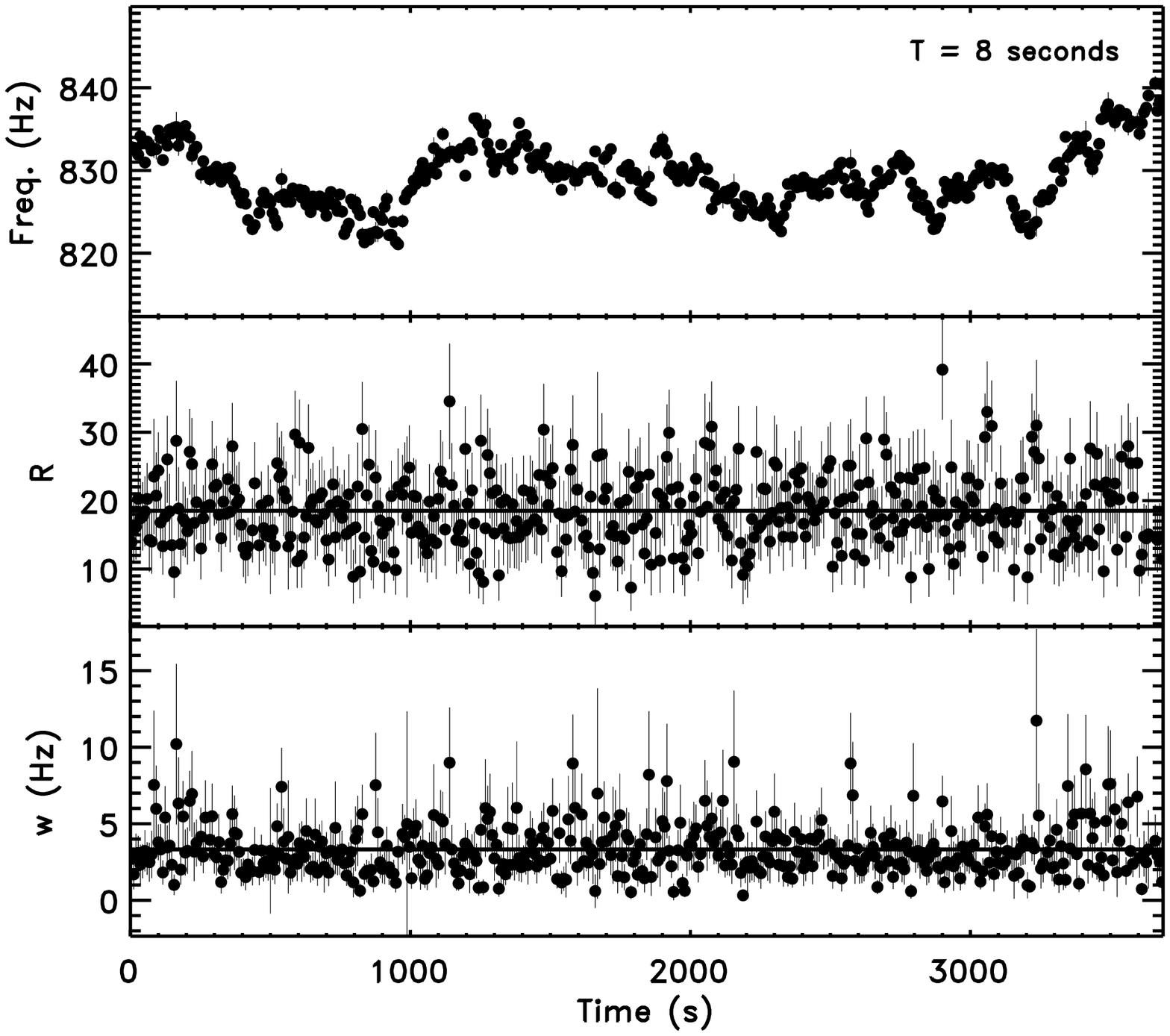}
\caption{Left) Dynamical PDS of 4U1608--522 as recorded during the second segment of the March 3rd, 1996 observations. The time resolution is 4 seconds. The X-ray light curve is shown at the top (in kcps), and the average PDS over the whole segment is shown in the vertical right panel (cut to a maximum of 2.7). A strong QPO is visible with a mean frequency around 830 Hz. Right) The MLEs for this segment of data, as derived from fitting one single PDS with an integration time of 8 seconds (from top to the bottom: the QPO frequency, $R$ and $w$). \label{bv11r_f6}}
\end{figure}

\4u\ is an X-ray transient and a prototypical high-frequency QPO source. The parameters of its high frequency QPOs have already been reported \citep{berger96,mendez98,barret05b,barret06}. Here we are interested in a subset of the data recorded by the PCA onboard {\it RXTE} on March 3rd, 1996 (ObsID 10072-05-01-00). This ObsID is split in three intervals. To produce the light curves, we have extracted all events in the detector channel range from 16 to 94. The smoothed dynamical PDS for the second interval of data is shown in Figure \ref{bv11r_f6}. As can be seen from that figure, both the frequency and amplitude of the QPO do not vary much during those observations. The MLE was applied by fitting one single periodogram of different integration times, down to 8 seconds. The results are shown in the right panel of Figure \ref{bv11r_f6} for an integration time of 8 seconds.  The mean significance of the excess power at the QPO frequency is about $\sim 8.7 \sigma$, and the mean ratio of $R/\sigma_R\sim 3.4$, indicating that the MLEs on $R$ and $w$ are not biased by more than $\sim 2-3$\% (see Figure \ref{bv11r_f2} for $T=8$ seconds). The QPO parameters are recovered for the first time on very short timescales, down to $8$ seconds from direct fitting. 

\begin{figure}[!t]
\epsscale{1.1250}
\plottwo{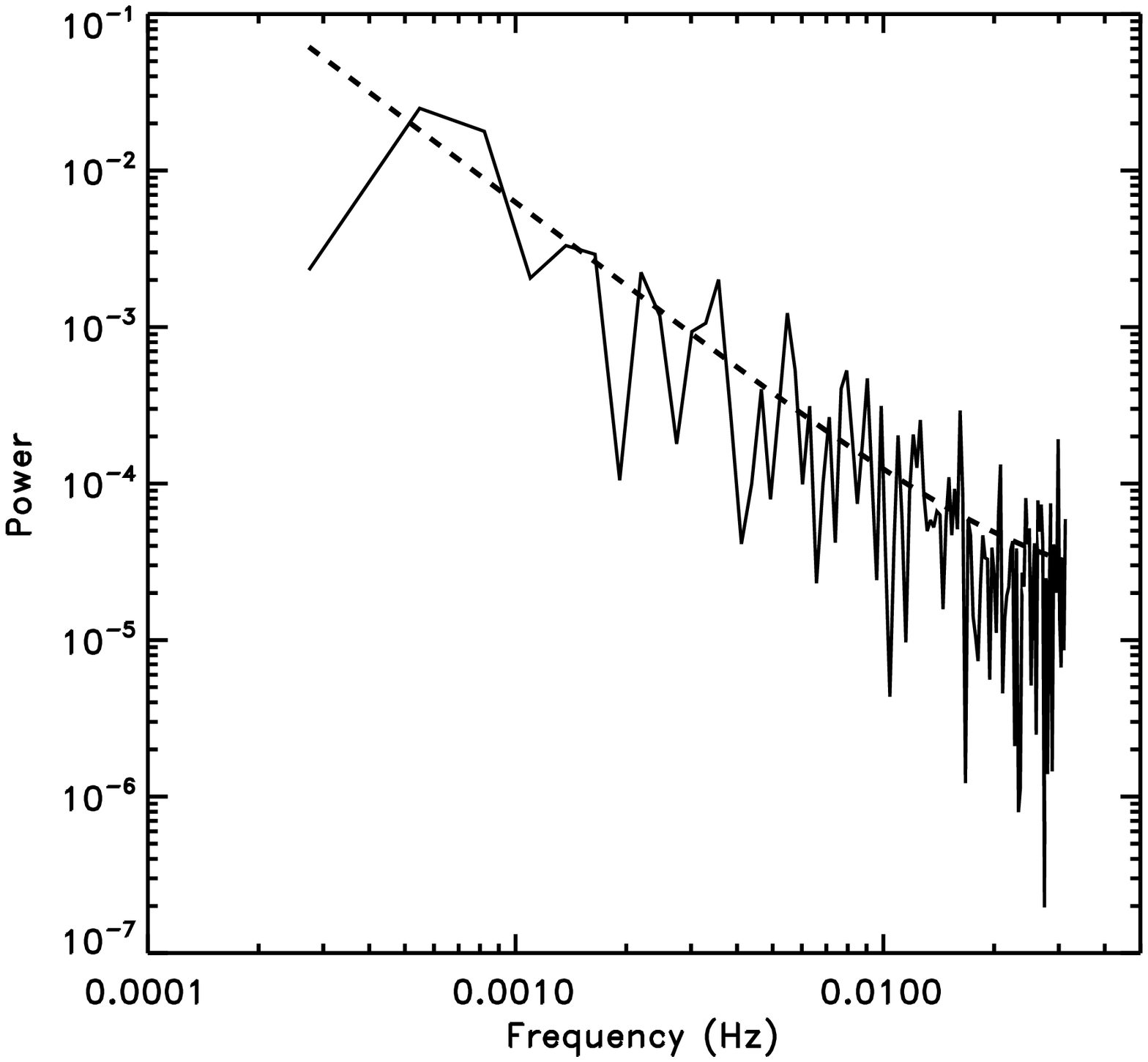}{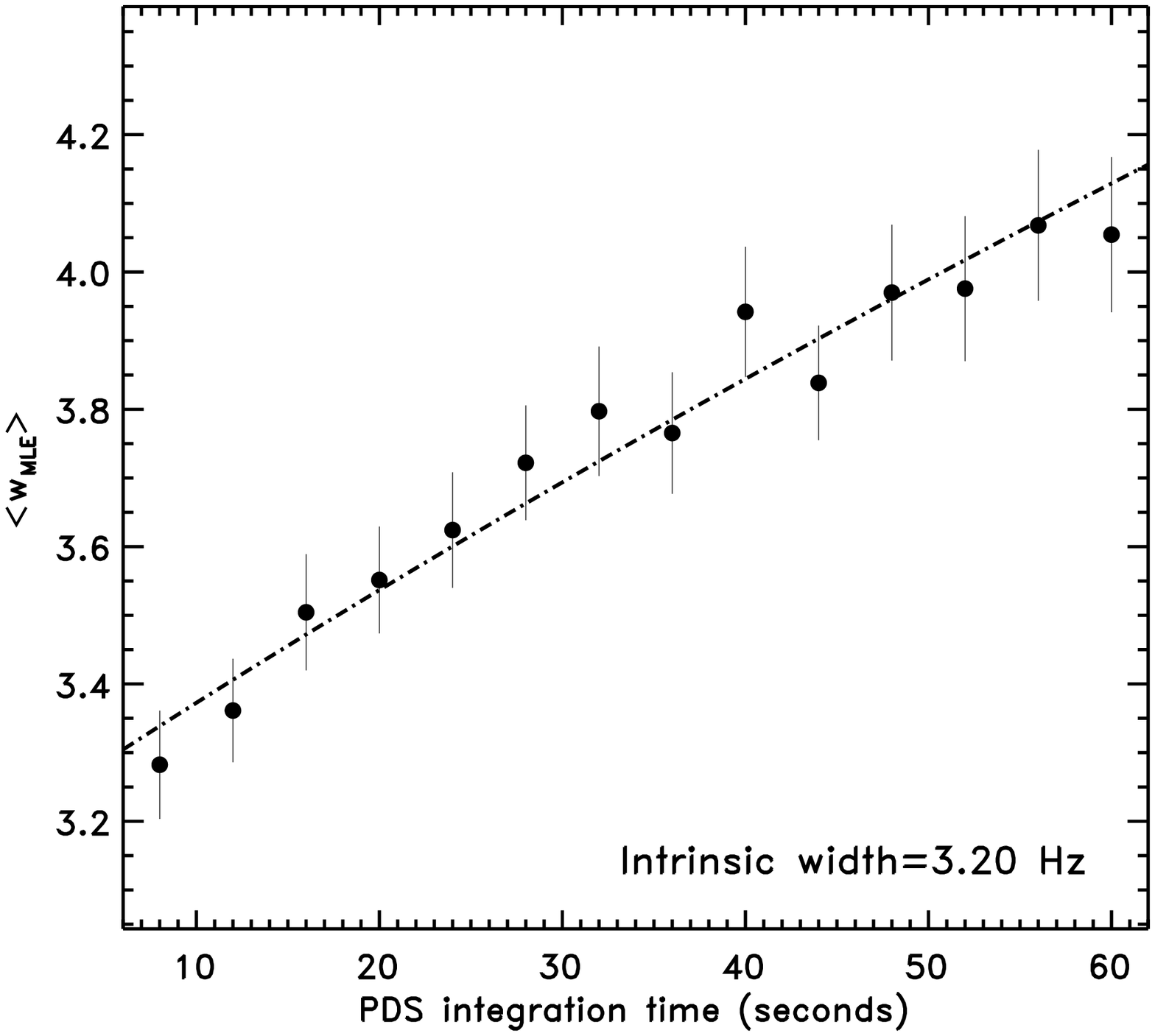}
\caption{Left) Periodogram of the frequency variations of \4u~as derived from the MLEs of the QPO frequency on a timescale of 16 seconds. The best power law fit is shown with a dashed line (index $1.8 \pm 0.2$). Right) Broadening of the QPO with increasing integration time of the fitted PDS. As expected from a random walk, the broadening goes like $\sqrt{T}$. The best fit is shown with a dot-dashed line. Such analysis enables us to derive the intrinsic QPO width to be 3.2 Hz, corresponding to a quality factor of $259 \pm 5$ at a frequency of $829$ Hz. \label{bv11r_f7}}
\end{figure}

In the left panel of Figure \ref{bv11r_f7} we show the power spectrum of the frequency variations, using the estimates of the QPO frequency on a timescale of 16 seconds. A maximum likelihood fit of this power spectrum indicates that the frequency evolution is consistent with a simple random walk, as previously hypothesized \citep{belloni05}. This power spectrum is adequately fitted by a simple model comprising a power law of index $1.8 \pm 0.2$ down to $\sim 1$ mHz, plus a constant that is consistent with that expected from the uncertainties on the frequency estimates.

Under the random walk hypothesis, we expect the broadening of the QPO due to the variable frequency to increase with $\sqrt{T}$. We have fitted single PDS with increasing $T$ and fitted the MLEs (corrected from the leakage bias) with a random walk model, assuming that the measured width for a given $T$ ($w_T$) includes a contribution from the intrinsic QPO width $w_{qpo}$ and a contribution from the drift ($w_{drift}$), such as $w_T^2=w_{qpo}^2+w_{drift}^2$, with $w_{drift} \propto \sqrt{T}$. As can be seen from Figure \ref{bv11r_f7}, a good fit is obtained when $T$ varies from 8 to 60 seconds. This enables to recover the intrinsic QPO width (in the limit of no frequency drift): $w_{QPO} = 3.20 \pm 0.03$ Hz. For a mean frequency of 829 Hz over the interval, this corresponds to a quality factor $Q=259 \pm 5$. Correcting for a positive residual MLE ($2$\%) bias (see Figure \ref{bv11r_f2}), this value goes up to $Q=264\pm6$, not significantly different. 

The mean of the MLEs of $w$ ($<w^{MLE}>=3.29\pm0.08$ Hz) measured for $T=8$ seconds (see Figure \ref{bv11r_f7}) is in fact already close to the intrinsic QPO width derived with the method above, and could therefore be used to approximate a mean $Q$ over a continuous observation (e.g. an ObsID in RXTE terminology). On the other hand, one should not take the weighted mean of $w^{MLE}$, as it underestimates the true width, due to the fact that smaller $w^{MLE}$ have smaller error bars (see Figure \ref{bv11r_f6}). Similarly, as discussed in Appendix C, shifting-and-adding PDS to a reference frequency (e.g. the mean frequency over an ObsID), and fitting the resulting averaged and shifted PDS leads also to a tiny negative bias on $w$ ($\sim 0.1-0.2$ Hz) and hence a slight overestimate of $Q$.

To conclude, using a simple random walk model and the MLE for different PDS integration times, we have derived a $Q$ factor of about 260, significantly larger than the value of 200 reported previously \citep{barret05b}. Unless the QPO shows rapid frequency variability below 8 seconds, the above value is likely to be close to the intrinsic QPO width. The higher quality factor derived, implying a longer the lifetime for the underlying oscillator, puts even even more stringent constraints on QPO models, as discussed in \cite{barret05b}. It also shows that the MLE, by enabling the unbiased estimation of QPO parameters on the shortest timescales permitted by the statistical quality of the data provides a clear improvement over previous methods. More data will be analyzed to test the random walk hypothesis, to measure the maximum $Q$ factor of kHz QPOs in neutron star systems, and to determine more accurately the dependency of the quality factor with frequency, in particular around the frequency, where $Q$ drops very rapidly: a feature that has been interpreted as a possible signature of the approach to the innermost stable circular orbit \citep{barret06,barret07}.

\section{Discussion and conclusions}

We have demonstrated the application of maximum likelihood fitting to periodogram data showing high frequency QPOs. The method is simple to implement and, importantly, remains valid in the non-Gaussian regime, e.g. when applied to raw periodogram data. This means it can be used in situations where averaging many periodograms is not desirable, e.g. when short time series are to be analysed at the highest frequency resolution. The simulation tests discussed in section $3$ show the method gives accurate and unbiased estimates of the QPO parameters, and can be used to probe QPO properties on timescales as short as $8$ seconds (for strong kHz QPOs). By contrast the standard \minchi2\ fitting method gives biased parameters with larger uncertainties when applied to such short sections of data because the spectral estimates are not Gaussian distributed. 

We have applied the MLE to an {\it RXTE} observation of the $\sim 830$ Hz QPO of the neutron star low-mass X-ray binary \4u. For one section of the observation we are able to recover the QPO frequency, width and amplitude on timescales down to $8$ seconds. We have shown that the frequency variations are consistent with a random walk in the observation analyzed, and we have inferred the intrinsic width of the QPO to correspond to a quality factor of $\sim 260$ at 830 Hz: a value significantly higher than previous estimates \citep{barret05b}. Of course, MLE fitting of time-resolved periodograms is not the only approach available for investigating the variability of the kHz QPOs. Another is to fit time-dependent models directly to the time-frequency data using hierarchical models \citep{gelman07}. This is beyond the scope of this paper whose aim is to highlight the improvement (in e.g. bias and variance of the fitted QPO parameters, and in some cases time resolution) that can be gained by using Whittle MLE rather than the $\min(\chi^2)$ method that is currently the standard procedure.

The method described here was applied for the first time to {\it RXTE} data. By enabling to measure the QPO parameters on very short timescales with negligible biases, this technique opens the way for new types of analysis to be performed on {\it RXTE} archive, as we have shown in the case of \4u\ for a limited subset of data. In most analysis so far, the QPO properties have been derived from averaging data sets, hence leading to a potential loss of information. The MLE does not require averaging. Hopefully, this will lead to a better understanding of the high frequency variability observed in accreting compact objects, both neutron stars and black holes. This is required to fully exploit the potential of this variability as a probe of strong gravity and dense matter \citep{klis00}.

\appendix
\section{Maximum likelihood fitting of averaged periodograms}

The periodogram is scattered around its expectation following a $\chi_2^2/2$ distribution. By averaging $M$ periodograms we find the average follows a $\chi_{2M}^2/2M$ distribution. From the properties of the chi-square distribution it is straightforward to derive the log likelihood formula in this more general case \citep[see also][]{appourchaux03}:
\begin{equation}
 \mathcal{S} = - 2 \ln \mathcal{L}
 = \nu \sum_{j=1}^{N-1} \left\{ \frac{I_j}{S_j} + \ln S_j + \left( \frac{2}{\nu} -1\right) \ln I_j + c(\nu) \right\}
\end{equation}
where $\nu = 2 M$ is the degrees of freedom of the relevant chi-square distribution, and $c(\nu)$ is a constant for fixed $\nu$ (not a function of $I_j$ or $S_j$). In the case of no averaging, i.e. $M=1$ ($\nu=2$), the above formula reduces to the usual log likelihood for the periodogram [since $c(\nu=2) = 0$].
The above formula may be used for maximum likelihood fitting of averaged periodograms with any $M$; in the limit of very large $M$ this is equivalent to the usual \minchi2\ method.

We have verified that fitting the average of $M$ PDS gives very similar results compared to fitting simultaneously $M$ PDS (minimizing the sum of $\mathcal{S}$ over $M$) (see Figures \ref{bv11r_f2} and \ref{bv11r_f3}). It is obviously as easy to implement, but runs much faster by a factor of $\sim M$. The errors on the MLEs can be computed the exact same way as in the case of $M=1$, e.g. directly from the Fisher matrix (see section \ref{sec2}).

\section{Spectral leakage}
The expectation value of the periodogram at Fourier frequency $f_j$ is
\begin{equation}
 E[I_j] = \int_{-f_N}^{+f_N} F(f_j-f^{\prime}) S(f^{\prime}) df^{\prime} = S(f) - b_I(f)
\end{equation}
where $S(f)$ is the true spectral density function, $F(f)$ is the Fejer kernel, and $f_N$ is the Nyquist frequency. The convolution by the Fejer kernel distorts the periodogram away from the true spectrum, and results from the finite sampling of the time series (see e.g. Brillinger 1975, chapter 5; Priestley 1981, chapter 6). The expected difference between the true spectrum and the periodogram is the bias on the periodogram: $b_I(f_j) = E[S(f_j) - I_j]$.

The Fejer kernel depends on the time series duration, $T$, and has a main lobe of width $\Delta f = 1/T$ plus oscillatory side-lobes that decay as $\sim 1/f^2$ either side. It is these side-lobes that cause spectral leakage -- the transfer of power between distant Fourier frequencies. Assuming a Lorentzian spectrum $S(f)$, typical of kHz QPOs, we can compute the effect of leakage by perfoming the above convolution numerically (on a fine grid of frequencies, $\delta f = 2^{-13}$). For well-resolved QPOs (i.e. $w > 1/T $) the convolved spectrum, i.e. the expectation of the periodogram, is very like a Lorentzian but with slightly larger width $w_C$. The bias on $w$ due to leakage, \bwl$ = w - w_C$, is itself a function of $T$. The convolution conserves the total power (integral under the curve) and so leakage does not directly bias $R$. However, a very small bias on $R$, almost always negligible, may be introduced by fitting a simple Lorentzian model to the distorted QPO profile, which can deviate slightly from a pure Lorentzian.

Figure \ref{bv11r_f8} shows this broadening effect, \bwl~ as a function of time series duration $T$, based on numerical calculation of the Fejer convolution. For longer time series the Fejer kernel becomes more concentrated and distorts the spectrum less, leading to smaller \bwl. For $T=4$ s the bias is $\approx 0.08$ Hz, within the precision of some estimates of QPO widths, whereas for $T$ longer than  $\sim 32$ s the bias is so small as to be practically insignificant.
\section{A small bias with the shift-and-add technique}
The so-called "shift and add" technique, as introduced by \citet{mendez98}, has been used extensively to study the power spectral properties of kHz QPOs. The method involves computing PDS estimates on short timescales (e.g. $T = 32$s), fitting for the QPO frequency, and using the best-fitting frequency to shift the frequency scale of each spectrum to one in which the QPO frequency is constant. This method can improve sensitivity to QPOs with rapidly variable frequencies that would get washed out in a straight average of the short timescale PDS estimates.

In the ideal case of the QPO frequency being known with zero error, the averaged spectrum will be broadened slightly by leakage due to the finite length of the time series intervals used to compute the short timescale PDS estimates (as discussed above and in Appendix B). But the method of shifting using fitted frequencies introduces another tiny bias to the estimation of QPO width. The fitted frequencies (that determine the shifts to be applied) for each interval are scattered around the true frequency without bias, but are closer to the centre of mass of each randomly sampled QPO spectrum than is the true frequency. On some occasions there will be more power above or below the true QPO frequency and these will contribute towards the profile of the QPO in the averaged spectrum. But if the data are shifted to reduce the spread of power around the fitted QPO location, this must result in less power in the tails and more power in the centre of the averaged QPO profile - i.e. a bias towards narrower QPOs in the average made using the shift and add technique. To estimate the magnitude of the  bias introduced by the shift-and-add, we used the fitted frequencies of the simulated data set generated above (i.e. taking 128 sets of 128 PDS with an integration time $T$ of 16 seconds). For each set, we have fitted the average PDS and the shifted-and-added PDS using \minchi2\ fitting. The estimates of $R$, $\nu_0$ and $a$ are consistent between the two sets of fits (straight average and shift-and-average), but the QPO width derived from the shifted-and-added PDS is significantly smaller than the one derived from the non-shifted averaged PDS by about $0.15$ Hz. This is the reason why we should use the sample mean of the estimates of $w^{MLE}$ to estimate the mean quality factor (section \ref{sec4}). The shift-and-add bias decreases with $T$ because the error on the fitted frequency decreases to zero: the width bias is less than $0.1$ Hz for $T = 24$ seconds. 

%
%
\acknowledgments{We wish to thank Thierry Appourchaux, Martin Boutelier, Dacheng Lin, Mariano Mendez and Cole Miller for useful comments on an earlier version of the manuscript. We are also grateful to an anonymous referee for a helpful report on the statistical aspects of the analysis presented in this paper. This research has made use of data obtained through the High Energy Astrophysics Science Archive Research Center Online Service, provided by the NASA/Goddard Space Flight Center.}
\bibliographystyle{apj} 

\newpage

%
%


\end{document}